\documentstyle{elsart}
                        
\begin{document}
\begin{frontmatter}
\title{Mass formulae and strange quark matter\thanksref{support}}
\thanks[support]{Supported in part by the National Natural Science
         Foundation of China under Grant No.\ 19575028.}
\author[ccast,nk]{G. X. Peng} \and
\author[nk,itp]{P. Z. Ning}
\address[ccast]{China Center of Advanced Science and Technology
                (World Lab.), \\ Beijing 100080, China }
\address[nk]{Department of Physics, Nankai University,
                               \\ Tianjin 300071, China }
\address[itp]{Institute of Theoretical Physics,
                  Academia Sinica, \\  Beijing 100080, China }

\begin{abstract}
    We have derived the popularly used parametrization formulae
 for quark masses at low densities and modified them at high densities
 within the mass-density-dependent model. The results are applied to
 investigate the lowest density for the possible existence of strange
 quark matter at zero temperature.
\end{abstract}

\begin{keyword}
  strange quark matter, mass formula, critical density.
\end{keyword}

\end{frontmatter}

{\bf PACS numbers:} 24.85.+p, 12.38.Mh, 12.39.Ba, 25.75.-q

\newpage

   The possible existence of strange quark matter (SQM) has been
 a focus of investigations~\cite{pen967} since Witten's
 conjecture~\cite{wit3084} that quark matter with strangeness per
 baryon of order unity may be bound. Because of the well-known
 difficulty of Quantum Chromodynamics in the nonperturbative domain,
 phenomenological models reflecting quark confinement are widely used
 in the study of hadron, and many of them have been successfully
 applied to investigate the stability and properties of SQM. One of
 the most famous models is the MIT bag model \cite{cho974} with which
 Farhi and Jaffe~\cite{far3084} find that SQM is absolutely stable
 around the normal nuclear density
  %  $n_0$ ($\approx 0.17$ MeV$\cdot$fm$^{-3}$)
 for a wide range of parameters.

     Another popularly used model is the mass-density-dependent model
in which quark confinement is achieved by requiring \cite{fow}
 \begin{equation}         \label{limitmq}
    \lim_{n_b\rightarrow 0} m = \infty,
 \end{equation}
where $m$ is the quark mass, $n_b$ is the baryon number density.
By using this model, Chakrabarty {\it et al}. \cite{cha22989} obtain a
significantly different result: only at very high densities does SQM
have the absolute stability.

   Benvenuto and Lugones \cite{ben} point out that this is the
consequence of an incorrect thermodynamical treatment. They add an
extra term to the energy expression, and get similar results to those
in the bag model.

  S. Chakrabarty {\it et al}. \cite{cha22989,cha5396} have already
discussed the limitation of the conventional MIT bag model which
assumes that the quarks are asymptotically free within the bag.
In order to incorporate the strong interactions between quarks, one
has to fall back on the perturbative theory, while the
mass-density-dependent model mimics not only the quark confinement,
but also the interactions between quarks. Therefore, the validity of
quark mass dependence on density is of utter importance.

However, the popularly used quark mass formulae \cite{fow,cha22989,ben}
are pure parameterizations without any real support from underlying
field theories up to now. Our motivation in writing this letter is to
try to derive the relation between the quark mass and density from more
fundamental principles on one hand, and to study the critical density
for SQM on the other hand. By ``critical density'' we mean in this paper
that only above the density does SQM have the possibility of existence.

   In a recent work \cite{pengrevc}, we have demonstrated that the quark
mass and quark condensates satisfy the following relation
\begin{equation}      \label{scale}
 \frac{m}{m_c}
      =\frac{1}{ 1-{\langle\bar{q}q\rangle_{n_b}}
                      /{\langle\bar{q}q\rangle_0}
                   },
\end{equation}
where we have suppressed the color index, the model parameter is changed
to $m_c$ from $m_{q0}$ which is saved for the original mass,
${\langle\bar{q}q\rangle}_0$ and $\langle\bar{q}q\rangle_{n_b}$ are the
quark condensates, respectively, in vacuum  and in strange quark matter
with baryon number density $n_b$. In this paper, we will see that
Eq.\ (\ref{scale}) has important properties of both confinement and
asymptotic freedom.

   According to the obvious equality
  \begin{equation}  \label{rqcl}
     \lim_{ n_b\rightarrow 0 }
    \frac { \langle\bar{q}q\rangle_{n_b} }
          { \langle\bar{q}q\rangle_0 }
            =1,
\end{equation}
we can expand the relative condensate as
\begin{equation}   \label{eqpen}
  \frac{ {\langle\bar{q}q\rangle}_{n_b} }
       { {\langle\bar{q}q\rangle}_0 }
   =1-\frac{n_b}
           { {\alpha^{\prime}} }
      +\mbox{higher orders in}\ n_b + \cdots,
\end{equation}
 where
\begin{equation}        \label{adef}
 \alpha^{\prime} = -\left(
 \frac{d}{dn_b}
               \frac{\langle\bar{q}q\rangle_{n_b}}
                    {\langle\bar{q}q\rangle_0}
                     \right)_{n_b=0}^{-1}.
\end{equation}

   At low densities, we can naturally ignore all terms in
Eq.\ (\ref{eqpen}) with orders in $n_b$ higher than 1 and
obtain
\begin{equation}   \label{appm}
  \frac{\langle\bar{q}q\rangle_{n_b}}
                         {\langle\bar{q}q\rangle_0}
  \approx 1 - \frac{n_b}{\alpha^\prime}.
\end{equation}

   Substituting Eq.\ (\ref{appm}) into Eq.\ (\ref{scale}), we get
 $m={m_c\alpha^\prime}/{n_b}$.

At high densities, the approximation (\ref{appm}) is no longer valid.
We will soon see that the quark mass is inversely proportional to
$n_b^{1/3}$, rather than just $n_b$.

   The Hellmann-Feynman theorem \cite{hel} indicates that
\begin{equation}      \label{hfe}
\langle\Psi(\lambda)|\frac{d}{d\lambda}H(\lambda)%
                    |\Psi(\lambda)\rangle
 =\frac{d}{d\lambda}\langle\Psi(\lambda)|H(\lambda)%
                                        |\Psi(\lambda)\rangle,
\end{equation}
where $H(\lambda)$ is a Hermitian operator, $|\Psi(\lambda)\rangle$
is the normalized eigenvector of $H(\lambda)$, $\lambda$\ is an
independent real parameter.

  We express the effective Hamiltonian density as
\begin{equation}
   H_{\mbox{\small eff}} =H^\prime\ + m \bar{q}q,
\end{equation}
where $H^\prime$\ is the kinetic  energy term.
In the quark mass-density-dependent model, the total (or effective)
mass depends on the density. It can be divided into two parts, namely,
$m=m_{0}+m_I$, where $m_0$ is the original mass independent of density,
and $m_I$ is the interacting mass mimicking the strong  interaction
between quarks.
Assuming that the quark condensate depends merely on the {\sl total}
quark number density, or, is blind to the configuration of the system,
we, for simplicity, consider only one flavour case in the following
derivation.

   Substituting $\int{d^3x}H_{\mbox{\small eff}}$ for $H(\lambda)$,
and $m_0$ for $\lambda$\ in Eq.\ (\ref{hfe}), we have
\begin{equation}
 \frac{dm}{dm_0} \langle\Psi|\int\!d^3\!x\bar{q}q|\Psi\rangle
   = \frac{d}{dm_0}\langle\Psi|
     \int\!d^3\!x{H}_{{\mbox{\small eff}}}|\Psi\rangle.
\end{equation}

  Because the expectation value of $H_{{\mbox{\small eff}}}$ should
equal that of $H_{{\mbox{\scriptsize QCD}}}$, we write down
\begin{equation}
 \frac{dm}{dm_0} \langle\Psi|\int\!d^3\!x\bar{q}q|\Psi\rangle
   = \frac{d}{dm_0}\langle\Psi|
     \int\!d^3\!x{H}_{{\mbox{\scriptsize QCD}}}|\Psi\rangle.
\end{equation}

    Applying this equation respectively to the cases
  $|\Psi\rangle=|n_b\rangle$ and $|\Psi\rangle=|0\rangle$,
  where $|n_b\rangle$ denotes the ground state of the quark matter
  at rest with baryon number density $n_b$, $|0\rangle$\
  is the vacuum, and then taking the difference, we obtain
\begin{equation}      \label{condif}
 \frac{\langle\bar{q}q\rangle_{n_b}}{\langle\bar{q}q\rangle_0}
  =1-\frac{1}{|\bar{q}q|_0}
  \frac{d\epsilon}{d{m_0}}/\frac{d{m}}{d{m_0}}
  =1-\frac{1}{|\bar{q}q|_0}\frac{\partial\epsilon}{\partial{m}},
\end{equation}
where the uniformity of the system has been taken into account,
 $|\bar{q}q|_0 \equiv -\langle\bar{q}q\rangle_0$,  $\epsilon$\ is
the energy density of the quark matter. Compared with the corresponding
formula for nuclear matter \cite{coh}, they are very similar in form.
But the physical contents are very different. The energy density there
depends on the nuclear density and the quark current mass which is an
independent quantity. Here we have the following chain relation

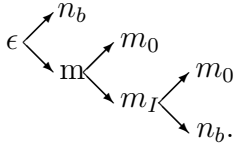
\setlength{\unitlength}{0.1cm}
\begin{picture}(60,25)
   \put(22,15){\makebox(1,2)[c]{$\epsilon$}}
\put(24,16){\vector(1,1){4}}
\put(24,16){\vector(1,-1){4}}
\put(30,19){\makebox(1,2)[c]{$n_b$}}
\put(30,11){\makebox(1,2)[c]{m}}
   \put(32,12){\vector(1,1){4}}
   \put(32,12){\vector(1,-1){4}}
   \put(39,15){\makebox(1,2)[c]{$m_0$}}
   \put(39,7){\makebox(1,2)[c]{$m_I$}}
\put(42,8){\vector(1,1){4}}
\put(42,8){\vector(1,-1){4}}
\put(49,11){\makebox(1,2)[c]{$m_0$}}
\put(49,3){\makebox(1,2)[c]{$n_b$.}}
\end{picture}

  This is why we should take $m_0$, rather than $m$, as the
substitute for $\lambda$.

   Combining Eqs.\ (\ref{scale}) and (\ref{condif}), we obtain
\begin{equation}   \label{geneq}
    m \frac{\partial\epsilon}{\partial{m}} % (1+\frac{d{m_I}}{d{m_0}})
     = {m_c|\bar{q}q|_0} \equiv B.
\end{equation}

At zero temperature, one has already known
\begin{equation}  \label{epsilon}
   \epsilon=\frac{gm^4}{16\pi^2} \left[x(2x^2+1) \sqrt{x^2+1}
   - \mbox{sh}^{-1}(x) \right],
 \end{equation}
 where sh$^{-1}(x)=\ln(x+\sqrt{x^2+1})$, $x=p_f/m$,
 the Fermi momenta $p_f=(\frac{18}{g}\pi^2 n_b)^{1/3}$,
 $g$ is the degeneracy factor(6 for quarks). If replacing the
 $\epsilon$\ in Eq.\ (\ref{condif}) by this expression, we can see
 that the relative condensate indeed has the similar expansion to
 Eq.\ (\ref{eqpen}).

  Substituting Eq.\ (\ref{epsilon}) into Eq. (\ref{geneq}), we have
\begin{equation}   \label{geneq0}
    x \sqrt{x^2+1} - \mbox{sh}^{-1}(x)
    =\frac{4\pi^2 B}{g m^4}.
 \end{equation}

   Generally, this equation can be solved numerically. However,
It can be analytically solved both at low and high densities.

   At low densities, we expand the left hand side of
Eq.\ (\ref{geneq0})  to $x^5$ term, and get
\begin{equation}  \label{mlow2}
   m=\frac{B}{3n_b}\left[\frac{1}{2}+\frac{1}{2}
    \left(
     1+\frac{2^{5/2}3^{13/3}\pi^{4/3}}{5g^{2/3}B^2} n_b^{8/3}
    \right)^{1/2}
                  \right].
\end{equation}

If we only expand Eq.\ (\ref{geneq0}) to $x^3$ term, or equivalently,
ignore the second term in the parentheses of Eq.\ (\ref{mlow2}),
then
 \begin{equation}  \label{mlow}
    m=\frac{B}{3n_b}.
 \end{equation}
This result was obtained from other arguments by the previous authors
\cite{fow} many years ago. It agrees to the confinement
property of quarks, and has been popularly applied to investigate the
properties of SQM \cite{cha22989,ben,cha5396}.

  At high densities, Eq.\ (\ref{geneq0}) becomes
 \begin{equation}   \label{eqapp}
    p_f^2 - m^2 \ln\left(\frac{2p_f}{m}\right)=\frac{4\pi^2 B}{g m^2}.
 \end{equation}

The first term on the left is much greater than the second one. So
we ignore the second term and get
 \begin{equation}  \label{mhigh}
    m=\frac{2\pi\sqrt{B/g}}{p_f}
     =\frac{(4\pi/9/\sqrt{g})^{1/3}\sqrt{B}}{n_b^{1/3}}
     \equiv \frac{\beta}{n_b^{1/3}}.
 \end{equation}

A more precise solution can be obtained by substituting this
expression for the $m's$\ in the second term on the left hand side
of Eq.\ (\ref{eqapp}):
 \begin{equation}  \label{mhigh2}
    m=\frac{2\pi\sqrt{B/g}}{p_f}\left[
    1+\frac{\pi^2B/g}{p_f^4}\ln(\frac{p_f^4}{\pi^2B/g})
                                \right].
 \end{equation}

Equation (\ref{mhigh}) or (\ref{mhigh2}) is in accordance with
the asymptotic freedom of QCD.

\begin{table}[htb]
\caption{Comparison of the relative errors(\%) for different mass
formulae. The densities are in unit of the normal nuclear density
$n_0=0.17$\ MeV$\cdot$fm$^{-3}$. Parameter $\beta$ is taken to be
(137 MeV)$^2$. ``*'' means over 100\%.}
                               \label{comparison}
 \begin{center}
\begin{tabular}{|c|r|r|r|r|r|r|r|}  \hline
 $n_b$ & 0.1 & 0.2 & 0.5 & 1. & 5.  &  10. & 15.    \\ \hline
Eq.(\protect\ref{mlow2})
& 0.00& 0.10  & 4.57 & 26.30 &  * & * & *  \\ \hline
Eq.(\protect\ref{mlow})
& 0.40  & 2.34  & 15.54 & 35.46 & 74.36 & 83.56 & 87.39   \\ \hline
Eq.(\protect\ref{mhigh})
& 72.31 & 56.90 & 31.34 & 16.71 & 3.23 & 1.53 & 0.98   \\ \hline
Eq.(\protect\ref{mhigh2})
& 86.74 & 43.09 & 10.75 & 1.66 & 0.44 & 0.23 & 0.15    \\ \hline
\end{tabular} \end{center} \end{table}

In Table \ref{comparison}, we show the relative errors for
different mass formulae. We see that the quark mass is inversely
proportional to density at very low densities. When the density
exceeds the nuclear density, the logarithmic term occurs, and the
quark mass changes asymptotically to be in reverse relation to
$n_b^{1/3}$. Therefore, we should apply Eq.\ (\ref{mhigh}) to the
investigation of SQM, namely,
\begin{eqnarray}
        &   m_{u,d}=\frac{\beta_0}{n_b^{1/3}},  &  \label{mud}  \\
        &   m_s   = \frac{\beta_s}{n_b^{1/3}}.  &  \label{ms}
 \end{eqnarray}

   The parameters $\beta_0$ and $\beta_s$ determine the critical
density $n_c$ rather simply
\begin{equation}    \label{ncexp}
    n_c = \frac{(\beta_s^2-\beta_0^2)^{3/4}}{\pi\sqrt{\gamma}},
\end{equation}
where $\gamma$ is dimensionless: $1.5<\gamma<2$. We use $\gamma=1.995$
in the calculations.

  To prove this expression,
  we assume the SQM to be a Fermi gas mixture of $u$, $d$, $s$ quarks
and electrons with chemical equilibrium maintained by the weak
interactions:
$
 d,s \leftrightarrow u+e+\overline{\nu}_{e}, \, \,
                                 s+u \leftrightarrow u+d.
$
   At zero temperature, the thermodynamic potential density of the
species $i$ is
\begin{equation}          \label{ztpd}
  \Omega_{i} = - \frac{g_{i}}{48\pi^2}
 \left[\mu_{i}(2\mu_{i}^2-5m_{i}^2)\sqrt{\mu_{i}^2-m_{i}^2}\right.
%                                            \nonumber      \\
% \mbox{}
 \left.+3m_{i}^{4}\ln\frac{\mu_{i}+\sqrt{\mu_{i}^{2}-m_{i}^{2}}}{m_{i}}
   \right],
\end{equation}
where $i=u, d, s, e$(ignore the contribution from neutrinos);
$g_i$ is the degeneracy factor with values $6$ and $2$ respectively
for quarks and for electrons; $m_{u,d}$ and $m_s$ to be replaced
by Eqs.\ (\ref{mud}) and (\ref{ms}).

   The corresponding baryon number density is
\begin{equation}
  n_i=-\frac{\partial\Omega_i}{\partial\mu_i}
     =\frac{g_i}{6\pi^2}(\mu_i^2-m_i^2)^{3/2}.
\end{equation}

   For a given $n_b$, the chemical potentials $\mu_i (i=u,d,s,e)$
 are determined by the following equations~\cite{far3084}
\begin{eqnarray}
   &  \mu_d  = \mu_s \equiv \mu,  &    \label{eqmu1}     \\
   & \mu_u + \mu_e = \mu,        &  \\
   & n_b=\frac{1}{3} (n_u + n_d + n_s), & \\
   & \frac{2}{3}n_u-\frac{1}{3}n_d-\frac{1}{3}n_s-n_e = 0.
                                      \label{eqmu4}   &
\end{eqnarray}

   The last two equations are equivalent to
\begin{eqnarray}
          & n_u-n_e=n_b,                  &    \\
          & n_d+n_s+n_e=2n_b.             &
\end{eqnarray}

   We therefore define a function of $\mu_e$
\begin{equation}
  F(\mu_e) = (\mu^2-m_{u,d}^2)^{3/2}+(\mu^2-m_s^2)^{3/2}
%                     \nonumber \\
             +\frac{1}{3}(\mu_e^2-m_e^2)^{3/2}-2\pi^2 n_b,
\end{equation}
where
\begin{equation}
  \mu=\mu_e+\sqrt{m_{u,d}^2+[\pi^2 n_b
           +\frac{1}{3}(\mu_e^2-m_e^2)^{3/2}]^{2/3}}.
\end{equation}

  Because $m_s>m_{u,d}$, the equation  $ F(\mu_e)=0 $
for $\mu_e$ has solution if and only if
\begin{eqnarray}
         \mu  & \geq & m_s,          \\
     F(\mu_e) & \leq & 0.
\end{eqnarray}

   At the critical density $n_c$, the equality signs in the above
should be taken, and so we can easily find
\begin{equation}       \label{nceq}
 1.5 \pi^2 n_c < (m_s^2-m_{u,d}^2)^{3/2} < 2 \pi^2 n_c.
\end{equation}

    Solving for $n_c$ from this inequality, we accordingly
 obtain Eq.\ (\ref{ncexp}).

    For a more exact $n_c$, we can numerically solve the following
 equation
\begin{eqnarray}
& m_s-\sqrt{m_{u,d}^2+[3\pi^2 n_c-(m_s^2-m_{u,d}^2)^{3/2}]^{2/3}} &
                                             \nonumber \\
& =\sqrt{m_e^2+\{3[2\pi^2 n_c-(m_s^2-m_{u,d}^2)^{3/2}]\}^{2/3}},  &
\end{eqnarray}
in the range
\begin{equation}
  \frac{(\beta_s^2-\beta_0^2)^{3/4}}{\sqrt{2}\pi}
        < n_c <
  \frac{(\beta_s^2-\beta_0^2)^{3/4}}{\sqrt{1.5}\pi}.
\end{equation}

    In fact, the expression (\ref{ncexp}) is precise enough for
practical applications because of the smallness of electron content.

 At $n_c$, the strangeness fraction becomes zero.
 When the density decreases further, the equation group
 (\ref{eqmu1}--\ref{eqmu4}) which determines the  configuration of
 the system has no solution. This indicates that $n_c$ is the lowest
 density for the possible existence of SQM.

    To calculate the critical density, we need to know the concrete
values of $\beta_0$ and $\beta_s$.

    Generally, there are two constraints on
$\beta_0$ \cite{far3084,cha22989,ben}: firstly, it should be so
big that the energy per baryon for light-flavour quark matter is
greater than 930 MeV in order not to contradict standard nuclear
physics; secondly, because we are interested in the possibility of
absolute stability of SQM, it should not exceed an upper limit so that
the energy per baryon can be less than 930 MeV for symmetric three
flavour quark matter. Therefore, $\beta_0$ must be in the range from a
fixed lower limit $\beta_{0min}$ to an upper limit $\beta_{0max}$.
By using the method in Ref.\ \cite{ben}, we obtain: $\beta_{0min}$ =
(135 MeV)$^2$, $\beta_{0max}$ = (147 MeV)$^2$.

   Because at very low densities the mass formula should be
Eq.\ (\ref{mlow}), we apply this formula to estimate the lowest
permissible value for $B$ independently by the same method. The result
is that $B$ should be no less than 69 MeV$\cdot$fm$^{-3}$. This
requires $\beta_0$ to be greater than (137 MeV)$^2$ according to
the relation $\beta =(4\pi/9/\sqrt{g})^{1/3}\sqrt{B}$
(see Eq.\ (\ref{mhigh})). It is remarkable that the requirement is
in accordance with the lower limit of $\beta_0$. With our purpose of
studying the lowest density for the possible existence of SQM in mind,
we take $\beta_0$ = (137 MeV)$^2$.

  As for $\beta_s$, there is no reliable method to fix it presently.
If regarding the nucleon as a special SQM with zero strangeness, one
has
\begin{equation}
\frac{\beta_s}{\beta_0}=\sqrt{\frac{B_s}{B}}
\sim \sqrt{\frac{2\sigma_{s}}{\sigma_N}} \approx 3.46,
\end{equation}
where the ratio $\sigma_s/\sigma_N$ has been taken to be 6 \cite{bro}.
The corresponding $n_c$ is $2.7 n_0$, not too far away from the
normal nuclear density. Therefore, wether SQM can exist near the
nuclear density is still an open question to study.

 \end{document}